\documentstyle[12pt,psfig]{article}
\newcommand{\nsect}{\setcounter{equation}{0}
\def\theequation{\thesection.\arabic{equation}}\section}

\newcommand{\appendixA}{\setcounter{equation}{0}
\def\theequation{\rm{A}.\arabic{equation}}\section*}
\newcommand{\appendixB}{\setcounter{equation}{0}
\def\theequation{\rm{B}.\arabic{equation}}\section*}

\catcode`\@=11
\def\marginnote#1{}
\def\ifmath#1{\relax\ifmmode #1\else $#1$\fi}

\def\msTop{m_{\,\widetilde{T}}}
\def\mstop{m_{\,\widetilde{t}}}

\def\bold#1{\setbox0=\hbox{$#1$}%
     \kern-.025em\copy0\kern-\wd0
     \kern.05em\copy0\kern-\wd0
     \kern-.025em\raise.0433em\box0 }

\def\GENITEM#1;#2{\par\vskip6pt \hangafter=0 \hangindent=#1
   \Textindent{$ #2$ }\ignorespaces}

\newcount\hour
\newcount\minute
\newtoks\amorpm
\hour=\time\divide\hour by60
\minute=\time{\multiply\hour by60 \global\advance\minute by-
\hour}
\edef\standardtime{{\ifnum\hour<12 \global\amorpm={am}%
    \else\global\amorpm={pm}\advance\hour by-12 \fi
    \ifnum\hour=0 \hour=12 \fi
    \number\hour:\ifnum\minute<100\fi\number\minute\the\amorpm}}
\edef\militarytime{\number\hour:\ifnum\minute<100\fi\number\minute}
\def\draftlabel#1{{\@bsphack\if@filesw {\let\thepage\relax
  \xdef\@gtempa{\write\@auxout{\string
    \newlabel{#1}{{\@currentlabel}{\thepage}}}}}\@gtempa
    \if@nobreak \ifvmode\nobreak\fi\fi\fi\@esphack}
     \gdef\@eqnlabel{#1}}
\def\@eqnlabel{}
\def\@vacuum{}
\def\draftmarginnote#1{\marginpar{\raggedright\scriptsize\tt#1}}
\def\draft{\oddsidemargin -.5truein
        \def\@oddfoot{\sl preliminary draft \hfil
        \rm\thepage\hfil\sl\today\quad\militarytime}
        \let\@evenfoot\@oddfoot \overfullrule 3pt
        \let\label=\draftlabel
        \let\marginnote=\draftmarginnote

\def\@eqnnum{(\theequation)\rlap{\kern\marginparsep\tt\@eqnlabel}%
\global\let\@eqnlabel\@vacuum}  }
\def\preprint{\twocolumn\sloppy\flushbottom\parindent 1em
        \leftmargini 2em\leftmarginv .5em\leftmarginvi .5em
        \oddsidemargin -.5in    \evensidemargin -.5in
        \columnsep 15mm \footheight 0pt
        \textwidth 250mmin      \topmargin  -.4in
        \headheight 12pt \topskip .4in
        \textheight 175mm
        \footskip 0pt

\def\@oddhead{\thepage\hfil\addtocounter{page}{1}\thepage}
        \let\@evenhead\@oddhead \def\@oddfoot{} \def\@evenfoot{}
}
\def\titlepage{\@restonecolfalse\if@twocolumn\@restonecoltrue\onecolumn
     \else \newpage \fi \thispagestyle{empty}\c@page\z@
        \def\thefootnote{\fnsymbol{footnote}} }
\def\endtitlepage{\if@restonecol\twocolumn \else  \fi
        \def\thefootnote{\arabic{footnote}}
        \setcounter{footnote}{0}}  
\catcode`@=12
\relax
\def\be{\begin{equation}}
\def\ee{\end{equation}}
\def\bea{\begin{eqnarray}}
\def\eea{\end{eqnarray}}
\def\simlt{\stackrel{<}{{}_\sim}}
\def\simgt{\stackrel{>}{{}_\sim}}

\def\NPB#1#2#3{{\it Nucl.~Phys.} {\bf{B#1}} (19#2) #3}
\def\PLB#1#2#3{{\it Phys.~Lett.} {\bf{B#1}} (19#2) #3}
\def\PRD#1#2#3{{\it Phys.~Rev.} {\bf{D#1}} (19#2) #3}
\def\PRL#1#2#3{{\it Phys.~Rev.~Lett.} {\bf{#1}} (19#2) #3}
\def\ZPC#1#2#3{{\it Z.~Phys.} {\bf C#1} (19#2) #3}

\def\MPLA#1#2#3{{\it Mod.~Phys.~Lett.} {\bf#1} (19#2) #3}

\def\AP#1#2#3{{\it Ann.~Phys.} {\bf#1} (19#2) #3}

\def\HPA#1#2#3{{\it Helv.~Phys.~Acta} {\bf#1} (19#2) #3}
\def\JETPL#1#2#3{{\it JETP~Lett.} {\bf#1} (19#2) #3}

\def\mst11{m_{\;\widetilde{t}_{1}}}

\def\st{\;\widetilde{t}}
\def\sT{\;\widetilde{T}}
\def\mw{m_W}
\def\mz{m_Z}
\def\mwbar{\overline{m}_{W_L}}

\def\mtop{m_t}
\def\mstop{m_{\st}}
\def\msTop{m_{\sT}}
\def\mstopbar{\overline{m}_{\st}}
\def\msTopbar{\overline{m}_{\sT}}
\def\mhiggs{m_h}
\def\mgoldstone{m_{\chi}}
\def\mhiggsbar{\overline{m}_h}
\def\mgoldstonebar{\overline{m}_{\chi}}
\def\dssv{{\cal D}_{SSV}}
\def\dsvv{{\cal D}_{SVV}}
\def\dvvv{{\cal D}_{VVV}}
\def\dsv{{\cal D}_{SV}}
\def\dvv{{\cal D}_{VV}}
\def\mg{m_g}
\def\mgp{m_{g'}}

\def\momegabar{\overline{m}_{\omega}}
\def\mrhobar{\overline{m}_{\rho}}

\def\mhbar{\overline{m}_{\cal H}}

\def\mst22{m_{\;\widetilde{t}_{2}}}
\def\mst12{m_{\;\widetilde{t}_{1,2}}}

\def\msb11{m_{\;\widetilde{b}_{1}}}
\def\msb22{m_{\;\widetilde{b}_{2}}}
\def\msb12{m_{\;\widetilde{b}_{1,2}}}

\def\mtilde2{\widetilde{m}^{2}}

\relax

\begin{document}
\topmargin-2.5cm
%
\begin{titlepage}
\begin{flushright}
FNAL-Pub-97/327-T\\
CERN-TH/97-190\\
IEM-FT-165/97 \\
hep--ph/9710401 \\
\end{flushright}
\vskip 0.3in
\begin{center}{\Large\bf Electroweak Baryogenesis and Higgs
\vskip .5 cm
and Stop Searches at LEP and the Tevatron}
\footnote{Work supported in part by the European Union
(contract CHRX/CT92-0004) and CICYT of Spain
(contract AEN95-0195).}
\vskip .5in
{\bf M. Carena~$^{a,b}$},
{\bf M. Quir\'os~$^{c}$} and {\bf C.E.M. Wagner~$^{a}$}
\vskip.35in
$^{a}$~CERN, TH Division, CH--1211 Geneva 23, Switzerland\\
$^b$~Fermi National Accelerator Laboratory, P.O. Box 500, Batavia, 
IL 60510, U.S.A.
$^{\ddagger}$~Instituto de Estructura de la Materia, CSIC, Serrano
123, 28006 Madrid, Spain
\end{center}
\vskip1.3cm
\begin{center}
{\bf Abstract}
\end{center}
\begin{quote}
It has been recently shown that the observed baryon number may originate
at the electroweak phase transition, provided that the 
Higgs boson and the lightest stop are sufficiently light. 
In this work, we perform a detailed analysis, including
all dominant two-loop finite temperature corrections to the 
Higgs effective potential, as well as the non-trivial effects proceeding
from the mixing in the stop sector,
to define the region of parameter
space for which electroweak baryogenesis can happen. 
The limits on the stop and Higgs
masses are obtained by taking into account 
the experimental bounds on these quantities, as well as 
those coming from the requirement of 
avoiding dangerous color breaking minima. We find for the
Higgs mass $m_h \simlt 105$~GeV, while the stop mass may be
close to  the present experimental bound and 
must be smaller than,
or of order of, the top quark mass.
These results provide a very strong motivation for further
non-perturbative analysis of the electroweak phase transition,
as well as for the search for Higgs and stop particles at
the LEP and Tevatron colliders.
\end{quote}
\vskip1.cm

\begin{flushleft}
October 1997 \\
\end{flushleft}
\end{titlepage}
\setcounter{footnote}{0}
\setcounter{page}{0}
\newpage
%
\nsect{Introduction}
Recently, there has been renewed interest in the idea of generating
the observed baryon number \cite{baryogenesis} 
at the electroweak phase transition \cite{anomaly}-\cite{first}.
A physical realization of this idea seems to be possible only
by going beyond the  Standard Model formulation. In fact, if no
fundamental standard model singlet scalar field is present in the
low energy effective theory, electroweak baryogenesis seems to 
be naturally induced only if the Higgs mass is sufficiently light, 
and if new light scalar particles, 
with strong couplings to the
Higgs field are present in the theory \cite{AndH}. Both effects tend to 
induce a strongly first order electroweak phase 
transition \cite{improvement}-\cite{nonpert}, which allows to 
preserve  the generated baryon number. Moreover, new CP-violating
phases, beyond those present in the Standard Model, seem to be necessary
to fuel the baryon number generation \cite{CPSM}.

Minimal low energy supersymmetric extensions of the Standard Model
fulfill the three above mentioned requirements. 
A light CP-even Higgs naturally appears 
in these models. Indeed, an upper bound, of order 125 GeV
can generally be set on the 
lightest CP-even Higgs mass \cite{Higgs}-\cite{CEQW}. 
The superpartners of
the top quark, whose coupling to the Higgs is of order
of the top quark Yukawa coupling, provide the additional scalars
which help in inducing a relatively strong first order 
phase transition \cite{early}-\cite{mariano2}. 
For this to happen, the lightest stop should have
a mass of order of, or smaller than, the top quark 
mass \cite{CQW}-\cite{JRB}. Moreover,
new CP-violating phases, induced by the supersymmetry 
breaking parameters associated with the left-right stop mixing,
are naturally present in these theories \cite{CPMSSM}. A very recent 
analysis shows that an explanation of the observed baryon
number of the universe requires CP-violating phases of 
order one and chargino and neutralino masses not much
larger than 200 GeV \cite{CQRVW}. Similar results were
obtained~\cite{CJK} by different analysis (see also 
Refs.~\cite{Iiro2,Toni2,Worah}). 

The above results present a strong case for the search of
Higgs and top-squark particles at LEP and the Tevatron 
colliders \cite{ReviewCW}. 
It is therefore  of the highest interest to define,
in the most accurate possible way, 
the exact region of parameter space consistent with electroweak
baryogenesis in the $m_h$--$m_{\st}$ plane. 
As has been first shown in Refs.~\cite{JoseR,JRB},
two loop effects are very important to 
define the correct boundary of this region. 
In Ref.~\cite{Schmidt}
it was suggested that, in the case of zero mixing in the stop 
sector, the phase transition can be sufficiently strong 
for Higgs masses as high as 100 GeV, if the stop mass is inside
a narrow band, of order 150 GeV.  

In this work, we define the allowed parameter space, 
including all dominant two-loop corrections to the Higgs
effective potential and allowing
a non-trivial mixing in the stop sector. We show that stop mixing
effects highly increase the region of stop masses consistent with
electroweak baryogenesis. Since light stops may induce 
color breaking minima, 
which should be taken into account
in the analysis \cite{CQW}, the finite temperature 
effective potential along the squark and Higgs directions is 
computed.  We identify the regions of parameter space in which 
the physical vacuum is stable within the thermal evolution of the
universe, as well as those in which a two-step phase transition,
via a color breaking minimum,  can take place. Finally, we 
also define the region of parameter space
in which the system is driven to the physical 
vacuum state at high temperatures, but this state  becomes 
unstable at low temperatures.  We define the boundaries which
separate the regions consistent with the different 
above mentioned possibilities and  show that, even in the case of absolute
stability of the physical vacuum, stops as light 
as $100$~--~$120$~GeV may
be consistent with electroweak baryogenesis, for Higgs masses in
the range 80--100~GeV.

In section 2 we summarize the most important properties of the
effective potential at finite temperature. The expressions of
the effective potentials for the stop and Higgs fields, 
with all dominant finite temperature
two-loop corrections included, are given in appendix A and B,
respectively.  In section 3 we
define the different regions of parameter space to be studied
in this article. In secion 4 we present our results. We reserve
section 5 for our conclusions.

\section{Finite temperature effective potential}

In order to preserve the generated baryon number after
the electroweak phase transition, a large vacuum
expectation value of the Higgs field at the critical
temperature is required \cite{first}
\begin{equation}
\frac{v(T_c)}{T_c} \simgt 1,
\label{vT}
\end{equation}
where $v = \langle\phi\rangle$ and $v(0) \simeq 246$ GeV.
The phase transition must be hence strongly first order.
The implications of the above bound may only be obtained
by a detailed knowledge of the electroweak phase transition.
In the Standard Model, it has been shown that
the two-loop improved effective potential leads to values
of the order parameter and the critical temperature in
good agreement with those obtained through a lattice  non-perturbative
analysis \cite{improvement,nonpert,Jansen}. 
In this case the requirement of Eq.~(\ref{vT})
demands a value of the Higgs mass which has been already
ruled out by experimental searches.
In this work, 
we shall present a two-loop perturbative
analysis for the case of the Minimal Supersymmetric Standard
Model (MSSM), in the expectation of a complete
numerical simulation of this model.

Following the methods developed in Ref.~\cite{twoloop}, 
we compute the four dimensional 
finite temperature effective potential
keeping all dominant two-loop finite temperature corrections
induced by the Standard Model gauge bosons as well as by
the light squarks and Higgs particles. As has been first
shown in Ref.~\cite{JoseR}, the two-loop corrections 
to the Higgs effective potential in the Minimal Supersymmetric
Model are dominated by loops involving the light stops and the
$SU(3)$ gauge bosons, and strongly depend on the QCD
coupling. 

Although we reserve the technical details of the effective
potential calculation for appendix A, we summarize here
the most important properties of the two-loop finite
temperature effective potential for the Higgs field. 
First of all, at one loop,
the dominant corrections to the effective potential come
from the gauge bosons, the top quark and its supersymmetric
partners. We shall work in the limit in which the left handed
stop is heavy, $m_Q \simgt 500$ GeV. In this limit,
the supersymmetric corrections to the precision electroweak
parameter $\Delta\rho$ become small and hence, this allows 
a good fit to the electroweak precision data coming from
LEP and SLD \footnote{A lighter left-handed stop may also be
consistent with precision electroweak measurements \cite{HERA}. 
However, this requires sizeable values of the stop mixing mass 
parameter, which tend to suppress the phase transition strength.}. 
Lower values of $m_Q$ make the phase transition
stronger and we are hence taking a conservative assumption
from the point of view of 
defining the region consistent with electroweak baryogenesis. The 
left handed stop decouples at finite temperature, but, at zero
temperature, it sets the scale of the Higgs masses 
as a function of $\tan\beta$. For 
right-handed stop masses below, or of order
of, the top quark mass, and for large values of the CP-odd 
Higgs mass, $m_A \gg M_Z$,
the one-loop improved Higgs
effective potential admits a high
temperature expansion,
\be
\label{potMSSM}
V_0(\phi) +V_1(\phi,T) = -\frac{m^2(T)}{2} \phi^2 - T \;
\left[E_{\rm SM}\; \phi^3 +  (2 N_c) \frac{\left(\mstop^{2} +
\Pi_{\st_R}(T)\right)^{3/2}}
{12 \pi} \right] + \frac{\lambda(T)}{8} \phi^4+\cdots
\ee
where
\be
\label{polarization}
\Pi_{\st_R}(T)=\frac{4}{9}g_s^2 T^2+
\frac{1}{6}h_t^2 \left[1+\sin^2\beta \left(1 - \widetilde{A}_t^2/m_Q^2
\right)\right]
T^2+\left(\frac{1}{3}-\frac{1}{18}|\cos 2\beta|\right)g'^2 T^2
\ee
is the finite temperature self-energy contribution to
the right-handed squarks (see appendix A), $g_s$ is the strong
gauge coupling, $N_c = 3$ is the number of colours and 
\be
\label{ESM}
E_{\rm SM}\sim \frac{1}{3}\
\left(\frac{2M_W^3+M_Z^3}{2 \pi v^3}\right). 
\ee
is the cubic term coefficient in the Standard Model case.

Within our approximation, the lightest stop mass is approximately
given by
\begin{equation}
\mstop^2 \simeq m_U^2 + 0.15 M_Z^2 \cos2\beta + m_t^2 
\left(1- \frac{\tilde{A}_t^2}{m_Q^2}\right)
\end{equation}
where $\tilde{A}_t = A_t - \mu/\tan\beta$ is the stop mixing 
parameter and $m_t = h_t \phi \sin\beta/\sqrt{2}$.
As was observed in Ref.~\cite{CQW}, the phase transition strength
is maximized for values of the soft breaking parameter 
$m_U^2 = - \Pi_{\st_R}(T)$, for which the coefficient of the cubic
term in the effective potential,  
\be
E \simeq E_{\rm SM}+ \frac{h_t^3 \sin^3\beta \left(1 -
\widetilde{A}_t^2/m_Q^2\right)^{3/2}}{4 \sqrt{2} \pi},
\label{totalE}
\ee
which governs the strength of
the phase transition,
\begin{equation}
\frac{v(T_c)}{T_c} \simeq \frac{4 E}{\lambda},
\end{equation}
can be one order of magnitude larger than $E_{\rm SM}$~\cite{CQW}.
In principle, the above would allow a sufficiently strong first order
phase transition for Higgs masses as large as 100 GeV.
However, it was also noticed that such large negative values of
$m_U^2$ may induce the presence of color breaking minima at
zero or finite temperature \cite{CW,CQW}. Demanding the absence
of such dangerous minima, the one loop analysis leads to 
an upper bound on the lightest CP-even
Higgs mass of order 80 GeV. This bound was obtained
for values of $\widetilde{m}_U^2 = - m_U^2$ of order (80 GeV$)^2$.

The most important two loop corrections are of the form
$\phi^2 \log(\phi)$ and, as said above, are induced by 
the Standard Model weak gauge bosons as well as by the stop and
gluon loops \cite{twoloop,JoseR}. 
It was recently noticed that the coefficient
of these terms can be efficiently obtained by the study of the
three dimensional running mass of the scalar top and Higgs fields
in the dimensionally reduced
theory at high temperatures \cite{Schmidt}. Equivalently, in a four
dimensional computation of the MSSM Higgs effective potential
with a heavy left-handed stop, we obtain
\begin{eqnarray}
V_2(\phi,T) & \simeq & \frac{\phi^2 T^2}{32 \pi^2}
\left[\frac{51}{16}g^2  - 3 \left[ h_t^2 \sin\beta^2 \left(1- 
\frac{\widetilde{A}_t^2}{m_Q^2}\right)\right]^2
+  8 g_s^2 h_t^2 \sin^2\beta \left(1- \frac{\widetilde{A}_t^2}{m_Q^2}\right)
\right] \log\left(\frac{\Lambda_H}{\phi}\right) 
\nonumber\\
\end{eqnarray}
where the first term comes from the Standard Model
gauge boson-loop contributions, 
while the second and third terms come from the
light supersymmetric particle loop contributions.
The scale 
$\Lambda_H$ depends on the finite corrections, which may be
obtained by the expressions given in appendix A \footnote{In the
numerical computations, we use the whole expression given in appendix A.}.
As mentioned above, the two-loop  
corrections are very important and, as has been shown in 
Ref.~\cite{JoseR}, they can make the phase transition strongly first
order even for $m_U \simeq 0$. 

An analogous situation occurs in the $U$-direction. The one-loop 
expression is approximately given by
\begin{equation}
V_{0}(U) + V_1(U,T)  = 
\left(-\widetilde{m}_U^2 + \gamma_U T^2 \right) U^2 -
T E_U U^3 + \frac{\lambda_U}{2} U^4,
\label{upot}
\end{equation}
where
\begin{eqnarray}
\gamma_U & \equiv & \frac{\Pi_{\st_R}(T)}{T^2} \simeq
\frac{4 g_s^2}{9} + \frac{h_t^2}{6}\left[ 1 +
\sin^2\beta (1 - \widetilde{A}_t^2/m_Q^2)
\right] ; \;\;\;\;
\lambda_U \simeq \frac{g_s^2}{3}
\nonumber\\
\label{totalEu}
E_U & \simeq &  \left[\frac{\sqrt{2} g_s^2}{6 \pi} \left( 1 +
\frac{2}{3\sqrt{3}} \right) \right] \\
& + &
\left\{ \frac{g_s^3}{12\pi}\left(\frac{5}{3\sqrt{3}} + 1\right)
+ \frac{h_t^3 \sin^3\beta (1-\widetilde{A}_t^2/m_Q^2)^{3/2}}{3 \pi} \right\}.
\nonumber
\end{eqnarray}
The
contribution to $E_U$ inside the squared brackets comes
from the transverse gluons, while the one inside the curly
brackets comes from the squark and Higgs contributions~\cite{CQW}.

Analogous to the case of the field $\phi$, the two loop corrections
to the $U$-potential are dominated by gluon and stop loops and are
approximately given by
\begin{equation}
V_2(U,T) = \frac{U^2 T^2}{16\pi^2}\left[  \frac{100}{9} 
g_s^4 - 2 h_t^2 \sin^2\beta \left(1-\frac{\widetilde{A}_t^2}{m_Q^2}
\right) \right] \log\left(\frac{\Lambda_U}{U}\right)
\end{equation}
where, as in the Higgs case, the scale $\Lambda_U$ may only
be obtained after the finite corrections to the effective
potential, given in appendix B, are computed.

Once the effective potential in the $\phi$ and $U$ directions are
computed, one can study the strength of the electroweak phase
transition, as well as the 
presence of potential color breaking
minima (see section 3). 
At one-loop, it was observed that requiring the 
stability of the physical vacuum at zero temperature was enough
to assure the absolute stability of the potential at finite temperature.
As has been first noticed in Ref.~\cite{Schmidt},
once two loop corrections are included, the situation is more 
complicated.
We shall proceed with this analysis in the next section.

\section{Color breaking minima}
 
As explained above, we work in the limit of large values
of $m_Q$, for which small contributions to $\Delta\rho$
are expected. As has been shown in Ref.~\cite{CQW}, in this limit
a color breaking minimum with $\langle Q\rangle \neq 0$ does not develop
at $T=0$, unless the stop mixing parameter $\widetilde{A}_t$
is very large. These large values of $\widetilde{A}_t$ suppress
the strength of the first  order phase transition and are
hence of no interest for this study. 
We expect no modification 
of this conclusion at $T \neq 0$, 
so far as the relation $m_Q \gg T$ is
preserved.

At zero temperature, for $Q = 0$ the minimization of the 
effective potential for the fields $\phi$ and $U$ shows
that the true minima are located for vanishing
values of one of the two fields. The two set of minima
are connected through a family of saddle points for
which both fields acquire non-vanishing  
values. Due to the nature of the high
temperature corrections, we do not expect a modification
of this conclusion at finite temperature. A detailed
analysis of this problem would however be needed to decide whether
minima with $\phi \neq 0$ and $U \neq 0$ for $Q = 0$
exist at finite temperature. We shall
restrict our analysis to the behaviour of the electroweak
and color breaking phase transitions in the directions
$\phi \neq 0$, $U =0$ and $U \neq 0$ and $\phi = 0$, respectively. 

Two parameters control the presence of color breaking minima:
$\widetilde{m}_U^c$, defined as the smallest value of $\widetilde{m}_U$
for which a color breaking minimum deeper than the electroweak
breaking minimum is present 
at $T = 0$, and $T_c^U$, the critical
temperature for the transition into a color breaking minimum in the
$U$-direction. The value of $\widetilde{m}_U^c$ may be obtained by
analysing the effective potential for the field $U$ at zero temperature,
and it is approximately given by \cite{CW,CQW}
\begin{equation}
\widetilde{m}_U^c \simeq
\left( \frac{m_H^2 \; v^2 \; g_s^2}{12} \right)^{1/4}.
\label{boundmu}
\end{equation}

Defining the critical temperature as that at which the potential
at the symmetry preserving and broken minima  are degenerate,
four situations can happen in the comparison of the critical
temperatures along the $\phi$ ($T_c$) and $U$ ($T_c^U$) transitions:

\begin{description}
\item[a)] $T_c^U < T_c$;   $\widetilde{m}_U < \widetilde{m}_U^c$ 
\item[b)] $T_c^U < T_c$;   $\widetilde{m}_U > \widetilde{m}_U^c$ 
\item[c)] $T_c^U > T_c$;   $\widetilde{m}_U < \widetilde{m}_U^c$ 
\item[d)] $T_c^U > T_c$;   $\widetilde{m}_U > \widetilde{m}_U^c$ 
\end{description}

In case a), as the universe cools down, a phase transition into
a color preserving minimum occurs, which remains stable until
$T = 0$. This situation, of absolute 
stability of the physical vacuum, is the most
conservative requirement to obtain electroweak baryogenesis.
We shall generally demand these conditions to define the allowed
parameter space in the $m_h$--$m_{\st}$ plane.
In case b), at $T = 0$ the color breaking minimum is deeper than
the physical one implying that
the color preserving minimum becomes unstable
for finite values of the temperature, with $T<T_c$. A physically
acceptable situation may only occur if the lifetime of the 
physical vacuum is smaller than the age of the universe. In
the following, we shall denote this situation 
as \lq\lq metastability''.
In case c), as the universe cools down, a color breaking minimum
develops which, however, becomes metastable as the temperature
approaches zero. A physically acceptable situation can only
take place if a two step phase transition occurs, that is
if the color breaking minimum has a lifetime lower than the 
age of the universe at some temperature $T < T_c$~\cite{Schmidt}. 
Finally, in case d) the color breaking
minimum is absolutely stable and hence, the situation becomes
physically unacceptable.

As said before, at one-loop, the condition of having a 
color preserving minimum at zero temperature, 
$\widetilde{m}_U < \widetilde{m}_U^c$,  
automatically
ensures that $T_c^U < T_c$. Hence, case c) never occurs
and a two step phase transition is not allowed. 
On the contrary, as we shall show below, when two-loop
corrections to the effective potential
of the Higgs and stop fields are included, 
cases b) or c) may occur depending mainly
on the  values of the
mixing in the stop sector and on $\tan\beta$. 

Figure 1 shows the region of parameter space consistent with
a sufficiently strong phase transition for four different
values of the stop mixing $\widetilde{A}_t$. At the left
of the solid line, $v/T \simgt 1$. Hence, for a given
value of 
$\widetilde{A}_t$ and $m_h$, the solid line gives the upper
bound on the stop mass consistent with the
preservation of the generated baryon number at the 
electroweak phase transition. 
Since $v/T$ is inversely proportional to $m_h^2$, lower
values of $m_{\st}$ (larger values of $E$) are
needed as the Higgs mass is increased. The dashed line
represents the values for which the critical temperature
of the $U$ field is equal to the critical temperature
for the electroweak phase transition. This means that 
$T_c > T_c^U$ ($T_c < T_c^U$) on the right (left) of the
dashed lines. 
Similarly, the short-dashed
line is defined by the values of 
the Higgs and stop mass parameters for which $\widetilde{m}_U =
\widetilde{m}_U^c$, i.e. $\widetilde{m}_U < \widetilde{m}_U^c$
($\widetilde{m}_U > \widetilde{m}_U^c$) on the right (left) of
the short-dashed lines. 
Observe that, for the same value of the
stop mass parameters, larger Higgs masses are associated
with larger values of $\tan\beta$. 

From  Fig.~1, it is clear that for low values of the
mixing, $\widetilde{A}_t \simlt 200$ GeV, 
case a) or c) may occur but, contrary to what
happens at one-loop, case b) is not realized. 
For the case of no mixing, this result
is in agreement with the analysis of \cite{Schmidt}.
The region of absolute stability of the physical
vacuum for $\widetilde{A}_t \simeq 0$
is bounded to values of the Higgs mass of order
95 GeV. There is a small region at the right of the
solid line,
in which a two-step phase transition may
take place, for values of the parameters which
would lead to $v/T < 1$ for $T = T_c$, but may
evolve to larger values at some $T < T_c$ at which the 
second of the two
step phase transition into the physical vacuum takes place.
This region disappears
for larger values of the stop mixing mass parameter. 
Quite generally,  depending on the value of $\widetilde{A}_t$,
the allowed region can be 
larger if a two step phase transition is considered,
although the upper bound on the Higgs mass is not
modified (see section 4).
For values of the mixing parameter $\widetilde{A}_t$ between
200 GeV and 300 GeV, both situations, cases b) and c) may
occur, depending on the value of $\tan\beta$.  
For large values of the stop mixing,
$\widetilde{A}_t > 300$ GeV,
a two-step phase transition does not take place. The region 
of parameter space located at the left of both the
dashed and short-dashed lines in the figure, corresponds to the
case d) and hence is ruled out.

\section{Higgs and Stop mass constraints}

To define the allowed region of space in the $m_h$--$m_{\widetilde{t}}$
plane, one has to join all the allowed regions for different
values of $\widetilde{A}_t$. As the value of $\widetilde{A}_t$ 
increases, the stop mass values tend to decrease, and hence
the lower bound on the stop mass is usually obtained for larger
values of $\widetilde{A}_t$. 
Since for larger values of $\widetilde{A}_t$, 
the phase transition becomes weaker, large values of the 
Higgs mass may only be obtained for low values of $\widetilde{A}_t$.
If we demand absolute stability,
the upper bound on $m_h$ decreases   
quite fast with $\widetilde{A}_t$, and the 
Higgs mass is constrained to be below 105, 95, 85 GeV for 
$\widetilde{A}_t = 300$, 500, 600 GeV, respectively.
Observe that, as becomes clear from Fig.~1, the same point in the
$m_h$--$m_{\widetilde{t}}$ plane may correspond to different 
values of the Higgs and stop parameters and hence
could lead to different physical situations.

Figure 2 shows the allowed region of parameter 
space in the $m_h$--$m_{\st}$ plane. The solid 
line demarks the region where $v/T \simgt 1$ and the physical vacuum
is absolutely stable. The right handed boundary is defined by the
points of $v/T = 1$ and $\widetilde{A}_t \simeq 0$ (values of
$\widetilde{A}_t \simgt 100$ GeV are necessary to reach the
largest values of $m_h$, see Fig.~1).  The tiny region 
of parameter space, 
shown in the upper left panel of Fig.~1, for which
a two step phase transition may occur for values of the
stop mass at the right of the solid line, $v(T_c)/T_c < 1$,
is not shown in the figure.
Observe that the
uppermost values of $m_h$ are approximately constant, independent
of $m_{\widetilde{t}}$ for a large region of parameter space. 
What constrains the Higgs mass in the region 
of $\widetilde{A}_t \simlt 300$ GeV is not the mechanism
of electroweak baryogenesis, but just the fact that, for the 
given values of $\widetilde{A}_t$ and $m_Q = 1$~TeV, the Higgs mass cannot
reach a larger bound. Slightly stronger (weaker) constraints
would be obtained for $m_Q = 500$ GeV (2 TeV).
Finally, the left boundary is defined by
values of $\widetilde{m}_U = \widetilde{m}_U^c$, for 
$\widetilde{A}_t \simgt 300$ GeV. As said before,
for this range of values of
the stop mixing, and as happens at the one-loop level, the condition
of stability at zero temperature ensures the stability at finite
temperature, as it becomes 
clear  by observing the evolution of the
different regions in Fig.~1.

All together, and even demanding absolute stability
of the physical vacuum, electroweak baryogenesis seems to work
for a wide region of Higgs and stop mass values. Higgs masses
between the present experimental limit, of about 75 GeV~\cite{Higgsbound}, 
and around
105~GeV are consistent with this scenario. Similarly, the running
stop mass may vary from values of order 165 GeV (of the same
order as the top quark mass one) and 100 GeV. 
Observe that, due to the influence
of the D-terms, values of $m_{\widetilde{t}} \simeq 165$ GeV,
$\widetilde{A}_t \simeq 0$ and $m_h \simeq 75$ GeV, are achieved for
small positive values of $m_U$ (of order 40 
GeV for the case represented in  Fig.~2). Also observe that
for lower values of $m_Q$ the phase transition may become
more strongly first order and slightly larger values of the
stop masses may be obtained.

The region  bounded by the thin dashed lines in
Fig.~2 corresponds to values of $m_h$~-~$m_{\st}$~for
which a two step phase transition can take place. 
Even though the effect of stop mixing greatly enlarges this 
region, we obtain from the present study that a two step
phase transition may occur only
in a narrow region of stop masses.
The exact location of this band depends slightly on the 
value of $m_Q$. For lower values of $m_Q$, it moves to larger
values of $m_{\tilde{t}}$. 
Finally, the allowed parameter space may be greatly increased
if we consider the condition of metastability. However, to 
analyse whether this may lead to a physically acceptable situation,
the lifetime of the finite temperature metastable vacuum should
be computed, something which is beyond the scope of this article.
 
\section{Conclusions}

In this article, we have computed the region of stop and
Higgs masses consistent with the preservation of the
baryon number generated at the electroweak phase transition.
We showed that, once two loop corrections are consistently
included in both the Higgs and the stop effective potentials,
the electroweak 
phase transition may occur at one or two steps, with the additional
possibility of a metastable physical vacuum state. Demanding
the condition of absolute stability of the physical vacuum
at all temperatures, we find that the phase transition 
may be sufficiently strong if the Higgs mass 
75 GeV $\simlt m_h \simlt$ 105 GeV and the lightest stop
mass 100 GeV $\simlt m_{\widetilde{t}} \simlt m_t$. The
allowed $m_{\st}$--$m_h$ region 
may be  enlarged if the condition of
metastability of the physical vacuum is considered, with
stop mass parameters which are somewhat lower than in the
case of an absolutely stable vacuum.  A two
step phase transition may only occur for values of the
Higgs and stop masses which are also consistent with absolute stability,
although the necessary stop mass parameters are slightly different.

Our results present a strong case for Higgs and stop searches
at LEP and the Tevatron colliders. Indeed, by the end of 
1999, LEP should be able to explore Higgs masses as high
as 100 GeV, testing almost the whole region consistent with
electroweak baryogenesis. An upgrade in energy, up to 
$\sqrt{s} \simeq 200$ GeV during the years 1999--2000 will allow
to explore the remaining region. If the Higgs is found at
LEP, the Tevatron shall be able to test a wide region of the
allowed stop masses during the next run, to start at the end of 
1999. The stop mass reach of the upgraded Tevatron
depends on the stop decay
channels, but the whole region of stop masses 
consistent with this scenario will be 
explored if the ultimate upgraded Tevatron
(TeV33)  becomes operative. 
Observe that, LEP is beginning
to test the lowest values of the stop mass \cite{Higgsbound},
which are compatible with a metastable electroweak vacuum
state.

Limits on the stop and Higgs mass parameters can  also be obtained
from electroweak precision measurements and rare decay processes.
As explained in section 2,  the parameter $\Delta\rho$ 
sets a lower bound on the left handed stop mass. Another
important measurement is the branching ratio 
${\rm BR}(b \rightarrow s \gamma)$. Indeed, it has been recently 
argued~\cite{CQRVW} that the baryon number generation at the
electroweak phase transition is suppressed for large values of
the CP-odd Higgs mass~\footnote{Observe, however, that both the
Higgs mass and the value of $v(T_c)/T_c$ vary only slightly with
$m_A$, for values of $m_A \simgt 250$ GeV. 
The results presented in this
article are hence expected to remain valid for these values of $m_A$.}. 
Lower values of the CP-odd Higgs mass, and
hence of the charged Higgs mass, tend to enhance the branching
ratio of this rare decay with respect to 
the Standard Model value \cite{bsgasusy}. Since the
Standard Model prediction  for 
${\rm BR}(b\rightarrow s \gamma)$ \cite{nolbsga} is more than one standard
deviation above   the
experimental value~\cite{bsgaexp}, 
a consistent result may only be obtained if
the stop-chargino contributions partially cancel  the charged
Higgs induced enhancement. For the values of the parameters presented
in Figure 1, and for a relatively light charged Higgs,
$m_{H^+} \simeq$~300~GeV (and hence $m_A \simeq$~300~GeV), 
this can only be achieved for moderate
values of the stop mixing parameter $\widetilde{A}_t$ 
(see ref. \cite{ReviewCW}). The precise bound depends on the
value of $\tan\beta$, $\mu$ and the charged Higgs mass, but
values of $\widetilde{A}_t \simlt 200$ GeV tend to be disfavored. 
Although this fact does not modify the upper bound on the Higgs
mass, it shows that the largest values of the stop masses, 
obtained for low values of $\widetilde{A}_t$ tend to be
disfavored (see Fig.~1).

Finally, it is important to stress that, since 
the above results
are obtained by using a two-loop finite temperature 
effective potential method, a non-perturbative analysis
will be necessary to support these
calculations, providing a definite  check 
of the region of  parameter space
consistent with electroweak baryogenesis.

\begin{center}
{\bf Acknowledgements}
\end{center} \noindent
M.C. and C.W. would like to thank the Aspen Center for Physics,
where part of this work has been done.
We would like to thank M. Shaposhnikov,
M. Laine and M. Schmidt for very interesting
discussions.

\appendixA{Appendix A}

In this appendix we present some technical details concerning the effective
potential at finite temperature along the $\phi$-direction, including the
leading two-loop corrections. We will consider the case where the  pseudoscalar
mass $m_A$ is much larger than the temperature $T$. In this case only the light
$CP$-even Higgs is coupled to the thermal bath, while the thermal distribution
functions of the heavy neutral $CP$-even, the neutral $CP$-odd and the charged 
Higgses are Boltzmann suppressed and decouple from the thermal bath.
We will work in the 't Hooft-Landau gauge and in the 
${\overline{\rm MS}}$-renormalization scheme.        

We will follow the notation and conventions of Ref.~\cite{AE}. Dimensional
regularization $(D=4-2\epsilon)$ is used to evaluate divergent integrals. Poles
in $1/\epsilon$ and terms depending on $\iota_{\epsilon}$ 
will be dropped since
they are canceled by counterterms~\cite{AE}. 
We will fix the ${\overline{\rm MS}}$-scale
$\overline\mu$ to the temperature $T$, and hence put to zero all terms in 
$\ln(\overline{\mu}/T)$, and all the couplings and fields will be 
considered at the scale $\overline{\mu}=T$: $g(T),\, \phi(T),\cdots$.
    
The states that give the most important contributions 
to the effective potential are the SM-particles:
electroweak gauge
bosons, the top quark, the Higgs and Goldstone bosons
($W$, $Z$, $\gamma$, $t$, $h$, $\chi$, respectively), as well as  
the left and right handed (third generation) scalar tops
($\st_L$, $\st_R$). The SM fields have tree-level masses:
\begin{eqnarray}
\mw^2 & = & \frac{1}{4}g^2 \phi^2 \nonumber\\
\mz^2 & = & \frac{1}{4}(g^2 + g'^2)\;\phi^2 \nonumber\\
\mtop^2 & = & \frac{1}{2}\sin\beta^2 h_t^2\; \phi^2 \nonumber\\
\mhiggs^2 & = & \frac{1}{8}(g^2+g'^2)\cos^2 2\beta\;(3\phi^2-v^2) \nonumber\\
\mgoldstone^2 & = & \frac{1}{8}(g^2+g'^2)\cos^2 2\beta\;(\phi^2-v^2) 
\label{SMmasses}
\end{eqnarray}
and degrees of freedom (longitudinal, $L$, and transversal, $T$,
for the gauge bosons)
\begin{eqnarray}
n_{W_L}& = & 2 \qquad n_{W_T}= 4 \nonumber \\
n_{Z_L}& = & 1 \qquad n_{Z_T}= 2 \nonumber \\
n_{\gamma_L}& = & 1 \qquad n_{\gamma_T}= 2 \nonumber \\
n_{h}& = & 1 \qquad n_{\chi}= 3 \nonumber \\
n_{t}&= &-12 \ .
\label{dofSM}
\end{eqnarray}
The stop fields ($\st_L,\st_R$), with degrees of freedom,
\begin{equation}
n_{\st_L}=n_{\st_R}=6
\label{dofMSSM}
\end{equation}
have a squared mass matrix given by:
\begin{equation}
{\cal M}_{\st}^2=\left(
\begin{array}{cc}
m_Q^2+m_t^2+\frac{1}{8}(g^2-g'^2)\cos 2\beta\; \phi^2 & m_t \widetilde{A}_t \\
m_t \widetilde{A}_t & m_U^2+m_t^2+\frac{1}{6}g'^2\cos 2\beta\; \phi^2 
\end{array}
\right)
\label{stopmatrix}
\end{equation}
where the $L-R$ mixing is defined as
\begin{equation}
\widetilde{A}_t=A_t-\mu/\tan\beta \ .
\label{mixing}
\end{equation}
The matrix (\ref{stopmatrix}) is diagonalized by the angle $\theta_t$ 
defined by
\begin{equation}
\sin 2\theta_t=
\frac{2 {\cal M}^2_{12}}{\sqrt{({\rm Tr} {\cal M}^2)^2-4\det{\cal
M}^2}} \qquad 
\cos 2\theta_t=\frac{{\cal M}^2_{11}-{\cal M}_{22}^2}
{\sqrt{({\rm Tr} {\cal M}^2)^2-4\det{\cal M}^2}}
\label{angulo}
\end{equation}
and we will call $\st$ and $\widetilde{T}$ 
the light and heavy eigenstates, and $\mstop$ and
$\msTop$ the corresponding eigenvalues.

We now define thermal masses for the longitudinal components of gauge bosons
($W_L, Z_L,\gamma_L$) and for all scalars ($h,\chi,\st_L,\st_R$) 
by means of the corresponding self-energies. We include in
the thermal bath the SM particles ($W$, $Z$, $\gamma$, $t$, $h$
and $\chi$), as well as the light stop $\st$, which plays a
prominent role in the strength of the phase transition, and
(light) charginos and neutralinos, which were proven to be
essential for the generation of the baryon asymmetry \cite{CQRVW}.
The self-energies, to leading order, are given by:
\begin{eqnarray}
\Pi_W & = & \frac{7}{3}g^2\; T^2
\nonumber \\
\Pi_B & = & \frac{22}{9}g'^2\; T^2  \nonumber \\
\Pi_h & = & \frac{1}{16}(g^2+g'^2)\cos^2 2\beta\; T^2
 + \frac{5}{16}g^2\; T^2+\frac{5}{48}g'^2\; T^2+
 \frac{1}{2}h_t^2 \sin^2\beta\; T^2
\nonumber\\
\Pi_{\chi} & = & \Pi_h \nonumber\\
\Pi_{\st_L} & = & \frac{1}{3} g_s^2\; T^2+\frac{5}{16}g^2\; T^2 
-\frac{1}{72}g'^2
\left(\frac{7}{6}+\cos 2\beta\right) T^2
+\frac{1}{12}h_t^2\left(2+\sin^2\beta\right) T^2 \nonumber \\
\Pi_{\st_R} & = & \frac{4}{9} g_s^2\; T^2+\frac{1}{18}\left(6+\cos
2\beta\right)g'^2\;
T^2+\frac{1}{6}h_t^2\left[1+\sin^2\beta\left(1-\frac{\widetilde{A}_t^2}
{m_Q^2}\right)\right] T^2 \ .
\label{pis}
\end{eqnarray}
The thermal masses are defined, for the SM bosons, as:
\begin{eqnarray}
\mwbar^2 & = & \mw^2 + \Pi_W \nonumber \\
\overline{m}^2_{Z_L,\gamma_L}& = & 
\frac{1}{2} \left[\frac{1}{4}(g^2+g'^2)\phi^2+\Pi_W+\Pi_B\right.\nonumber \\
&\pm & \left.\sqrt{\left( (g^2-g'^2)\frac{\phi^2}{4}+\Pi_W-\Pi_B\right)^2
+\frac{1}{4} g^2 g'^2 \phi^4}\; \right]
\nonumber\\
\mhiggsbar^2 & = & \mhiggs^2+ \Pi_h \nonumber \\
\mgoldstonebar^2 & = & \mgoldstone^2 + \Pi_{\chi}
\label{masasbar}
\end{eqnarray}
and for the stops as:
\begin{equation}
\overline{{\cal M}}_{\st}^2=\left(
\begin{array}{cc}
m_Q^2+m_t^2+\frac{1}{8}(g^2-g'^2)\cos 2\beta\; \phi^2 + \Pi_{\st_L} 
& m_t \widetilde{A}_t \\
m_t \widetilde{A}_t & m_U^2+m_t^2+\frac{1}{6}g'^2\cos 2\beta\; \phi^2 
+\Pi_{\st_R} 
\end{array}
\right)\ .
\label{stopmatbar}
\end{equation}
Similarly one can diagonalize the mass matrix (\ref{stopmatbar}) by means of
the angle $\overline{\theta_t}$,
leading to the mass eigenstates $\mstopbar$ and $\msTopbar$. For $m_Q\gg T$: 
\begin{equation}
\sin\theta_t\simeq \sin\overline{\theta}_t \simeq
\frac{m_t\widetilde{A}_t}{m_Q^2}\ .
\label{angapph}
\end{equation}

We can now expand the effective potential as a sum
\begin{equation}
V(\phi,T)=V_0+V_1+V_2+\cdots
\label{potphi}
\end{equation}
where $V_n$ indicates the $n$-th 
loop potential in the resummed theory at finite
temperature. 

The tree-level potential can be written as:
\begin{equation}
V_0=-\frac{1}{2}m^2(\bar{\mu}) 
\phi^2+\frac{1}{32}(g^2+g'^2)\cos^2 2\beta \; \phi^4
\label{v0higgs}
\end{equation}
where
\begin{equation}
m^2(\bar{\mu}) = \frac{1}{2}m_Z^2(v) \cos^2 2\beta+
\sum_i \frac{n_i}{16\pi^2} m_i^2(v) \frac{d m_i^2(v)}{d v^2}
\left[\log(m_i^2(v)/\bar{\mu}^2)+\frac{1}{2}-C_i\right]
\label{m2T}
\end{equation}
where $i=W,Z,h,\chi,t,\st,\sT$, the constants $C_i$ are 
$C_i=5/6$ ($C_i=3/2$) for gauge bosons 
(scalar bosons and fermions), and the VEV of the Higgs field is normalized to
$v=246$ GeV.

We will perform daisy resummation on the $n=0$ modes of the
longitudinal components of the gauge bosons 
$W_L, Z_L,\gamma_L$ and for all modes of the
bosons $h$, $\chi$ and the light stop $\st$. No resummation is performed on
fermions (i.e. the top quark) or for heavy bosons which decouple from the
thermal bath (in particular the heavy stop $\sT$). With the above prescription
we can write the one-loop effective potential $V_1$ as:
\begin{equation}
V_1=\sum_{i} \frac{n_i}{64\pi^2}M_i^4\left(
\log \frac{M_i^2}{\bar{\mu}^2}-C_i\right)+\sum_i\frac{n_i}{2\pi^2}J^{(i)} T^4 
\label{1loopot}
\end{equation}
where, as said before, for the analysis of the phase transition, we shall take
$\bar{\mu} = T$. In the above, 
the sum extends over $i=W_L,Z_L,\gamma_L,W_T,Z_T,t,h,\chi,
\widetilde{t},\widetilde{T}$, and the masses $M_i$ are defined by
\begin{equation}
M_i=
\left\{
\begin{array}{ll}
m_i & i=W_L,Z_L,\gamma_L,W_T,Z_T,t,\widetilde{T} \\ & \\
\overline{m}_i & i=h,\chi,\widetilde{t} \ .
\end{array}
\right.
\label{masash}
\end{equation}
The thermal contributions $J^{(i)}$ are defined by
\begin{equation}
J^{(i)}=
\left\{
\begin{array}{ll}
{\displaystyle J_B(m_i^2)-\frac{\pi}{6}\left(\overline{m}_i^3-m_i^3\right)} &
i=W_L,Z_L,\gamma_L \\ & \\
J_B(m_i^2) & i=W_T,Z_T,\widetilde{T} \\ & \\
J_B(\overline{m}_i^2) & i=h,\chi,\widetilde{t} \\ & \\
J_F(m_i^2) & i=t \ ,
\end{array}
\right.
\label{jotas}
\end{equation}
and the thermal integrals $J_{B,F}$ are:
\begin{equation}
J_{B,F}(y^2)=\int_0^\infty dx\; x^2 \log\left(1\mp
e^{-\sqrt{x^2+y^2}}\right)\ . 
\label{termicas}
\end{equation}

The two-loop potential can be split into the SM ($V_2^{\rm SM}$)
and the MSSM ($V_2^{\rm MSSM}$) terms. 
\begin{equation}
V_2=V_2^{\rm SM}+V_2^{\rm MSSM}
\label{dos}
\end{equation}
The value of $V_2^{\rm SM}$ has been computed in Ref.~\cite{AE} 
where the full expression can be found. 
It is dominated by logarithmic terms that can be written as:
\begin{eqnarray}
V_2^{\rm SM}&=&\frac{g^2}{16\pi^2}
T^2\left[-3\left(\frac{3}{2}\mw^2+2\Pi_W\right)
\log\frac{\mw+2\mwbar}{3T} -\frac{63}{8}\mw^2\log\frac{\mw}{T}\right.
\nonumber\\
&+&\left. \frac{3}{4} \mw^2\log\frac{2\mwbar}{3T}+
\frac{3}{2}\mw^2\log\frac{2\mw}{3T}+3\mwbar\mw \right]
\label{dosSM}
\end{eqnarray}
where the approximation $g'=0$ has been used. There are also
non-logarithmic contributions involving the strong $g_s$ and
Yukawa $h_t$ couplings which are less important for the phase
transition. 

In the two-loop potential $V_2^{\rm MSSM}$ the strong and top
Yukawa couplings are involved in logarithmic contributions.
Therefore  we have found to be an excellent approximation to keep $g_s$
and $h_t$, and to neglect the weak couplings $g$ and $g'$. In
this approximation the relevant diagrams that 
contribute to $V_2^{\rm MSSM}$ are of two kinds:
sunset diagrams, labeled by $V_{XYZ}$, where $X$, $Y$ and $Z$ are the
propagating fields, and figure eight diagrams, labeled by $V_{XY}$, 
with propagating $X$ and $Y$ fields.
With this prescription $V_2^{\rm MSSM}$ can be decomposed as:
\begin{equation}
V_2^{\rm MSSM}=V_{\st\st g}+V_{\st\st h}+V_{g\st}+V_{\st h}+
V_{\st\chi}+V_{\st\st}\ ,
\label{dosloops}
\end{equation}
where $g$ stands for gluons. The different contributions are given by:
\begin{eqnarray}
V_{\st\st\; g}& = & 
-\frac{g_s^2}{4}(N_c^2-1){\cal D}_{SSV}(\mstopbar,\mstopbar,0)
\nonumber \\ 
V_{\st\st\; h}& = & -\left(
h_t^2\sin^2\beta\;\frac{\phi}{\sqrt{2}}+h_t \widetilde{A}_t\sin\beta
\sin\overline{\theta}_t\cos\overline{\theta}_t
\right)^2N_c\; H(\mhiggsbar,\mstopbar,\mstopbar)\; T^2 \nonumber\\
V_{g\st}& = &-\frac{g_s^2}{4} (N_c^2-1) {\cal D}_{SV}(\mstopbar,0) 
\nonumber \\ 
V_{\st\; h}& = & \frac{1}{2}h_t^2\sin^2\beta\; N_c\; 
I(\mstopbar)\,I(\mhiggsbar)\; 
\nonumber \\
V_{\st\;\chi}& = & \frac{3}{2}h_t^2\sin^2\beta\; N_c\; 
I(\mstopbar)\,I(\mgoldstonebar)\; 
\nonumber \\
V_{\st\st}& = & \frac{g_s^2}{6}N_c(N_c+1)\; I^2(\mstopbar)\; 
\label{2looph}
\end{eqnarray}
The functions involved in (\ref{2looph}) are all defined in Ref.
\cite{AE}. We can use them with the following prescriptions:
\begin{itemize}
\item
Cancel all poles in $1/\epsilon$, and $\iota_\epsilon$-terms,
with counterterms.
\item
Take the limit $\epsilon\rightarrow 0$ and keep all finite
contributions. 
\item
Cancel all $c_B$ terms for scalar bosons with thermal
counterterms. 
\item
Linear terms as $M_i T^3$ which appear in figure-eight diagrams
are canceled by thermal counterterms.
\end{itemize}

Following these rules, we obtain results which are in good
agreement with those found in Ref. \cite{JoseR}.

\appendixB{Appendix B}

In this appendix we will specify the effective potential at
finite temperature in the background field
$U\equiv\st_R^\alpha u_\alpha$, where $u_\alpha$ is a constant unit
vector in color space~\footnote{Using the $SU_c(3)$ invariance the vector 
$u_\alpha$ can be chosen as $u_\alpha=(1,0,0)$.},
which breaks $SU_c(3)$ into $SU(2)$. We will proceed
as in appendix A and present the result of the two-loop
calculation. We will neglect $g'$ and keep the strong $g_s$ and top Yukawa
$h_t$ couplings. In the following, we use $g'$ to denote one of the gluon
states appearing in the spectrum.

The states contributing to the effective potential \cite{CQW} are the 
four gluons $g$ and the gluon $g'$, five real
squarks $\omega$ (would-be goldstones) and the real squark $\rho$,
four light ${\cal H}$ and four heavy ${\cal Q}$ scalars, 
coming from the mixing between 
the SM Higgs doublet $H$ and the left-handed (third generation) squark doublet
$\widetilde{q}\equiv\widetilde{q}^\alpha u_\alpha$, 
and two degenerate massive Dirac fermions $f$ coming from the mixing between 
the left-handed (third generation) fermion doublet 
$q_L\equiv q_L^\alpha u_\alpha$ 
and the higgsino~\cite{CQW,Schmidt}. 
The masses of $g$, $g'$, $\omega$, $\rho$ 
and $f$ are given by:
\begin{eqnarray}
m_g^2 & = & \frac{1}{2}g_s^2 U^2 \nonumber\\
m_{g'}^2 & = & \frac{2}{3}g_s^2 U^2 \nonumber\\
m_{\omega}^2 & = & m_U^2+\frac{1}{3}g_s^2 U^2 \nonumber\\
m_{\rho}^2 & = & m_U^2+ g_s^2 U^2 \nonumber\\
m_f^2 & = & \mu^2+h_t^2 U^2 
\label{Umasses}
\end{eqnarray}
and their degrees of freedom are
\begin{eqnarray}
n_{g_L}& = & 4 \qquad n_{g_T}= 8 \nonumber \\
n_{g'_L}& = & 1 \qquad n_{g'_T}= 2 \nonumber \\
n_{\omega}& = & 5 \qquad n_{\rho}= 1 \nonumber \\
n_{f}&= &-8 \ .
\label{dofU}
\end{eqnarray}
The scalars ($H,\widetilde{q}$), with degrees of freedom
\begin{equation}
n_{H} = 4 \qquad n_{\widetilde{q}}= 4 
\label{otros}
\end{equation}
have a squared mass matrix
\begin{equation}
{\cal M}_{H,\widetilde{q}}^2=\left(
\begin{array}{cc}
-\frac{1}{2}m_h^2 + h_t^2\sin^2\beta\; U^2  & 
h_t\sin\beta\; U \widetilde{A}_t \\
h_t\sin\beta\; U \widetilde{A}_t & 
m_Q^2+ h_t^2 U^2  
\end{array}
\right)\ .
\label{Hqmatrix}
\end{equation}
After diagonalization of the matrix (\ref{Hqmatrix}) 
by the rotation defined through
the angle $\theta_{\cal H}$, defined as in (\ref{angulo}), we obtain the light 
${\cal H}$ and heavy ${\cal Q}$ eigenstates and the corresponding eigenvalues,
$m_{\cal H}^2$ and $m_{\cal Q}^2$.

Using the self-energies (\ref{pis}), and the gluon self-energy 
\begin{equation}
\Pi_g=\frac{8}{3}g_s^2 T^2
\label{pig}
\end{equation}
we can define the thermal masses for the longitudinal components
of gauge bosons ($g_L$, $g'_L$) and for all scalar bosons
($\omega$, $\rho$, ${\cal H}$, ${\cal Q}$) as:
\begin{eqnarray}
\overline{m}_{g_L}^2 & = & m_{g_L}^2+\Pi_g \nonumber\\
\overline{m}_{g'_L}^2 & = & m_{g'_L}^2+\Pi_g \nonumber\\
\overline{m}_{\omega}^2 & = & m_{\omega}^2 +\Pi_{\st_R} \nonumber\\
\overline{m}_{\rho}^2 & = & m_{\rho}^2 +\Pi_{\st_R} \nonumber\\
\end{eqnarray}
while $\overline{m}_{\cal H}^2$ and $\overline{m}_{\cal Q}^2$ are the 
eigenvalues of the thermally corrected squared mass matrix
\begin{equation}
\overline{{\cal M}}_{H,\widetilde{q}}^2=\left(
\begin{array}{cc}
-\frac{1}{2}m_h^2 + h_t^2\sin^2\beta\; U^2 +\Pi_{h} & 
h_t\sin\beta\; U \widetilde{A}_t \\
h_t\sin\beta\; U \widetilde{A}_t & 
m_Q^2+ h_t^2 U^2  +\Pi_{\st_L}
\end{array}
\right)\ .
\label{HqmatrixT}
\end{equation}
diagonalized by the rotation $\overline{\theta}_{\cal H}$. For $m_Q\gg T$:
\begin{equation}
\sin\theta_{\cal H}\simeq\sin\overline{\theta}_{\cal H}\simeq
\frac{h_t\sin\beta\; U\widetilde{A}_t}{m_Q^2}\ .
\label{angappU}
\end{equation}

We will now expand the effective potential in a loop expansion as in
(\ref{potphi}), 
\begin{equation}
V(U,T) = V_0 + V_1 + V_2 +...
\end{equation} 
The tree level potential can be written as:
\begin{equation}
V_0=m_U^2(\bar{\mu})\; U^2+\frac{1}{6}g_s^2\; U^4
\label{VoU}
\end{equation}
with
\begin{equation}
m^2_U(\bar{\mu})=m_U^2 -
\sum_i \frac{n_i}{32\pi^2} m_i^2(u) \frac{d m_i^2(u)}{d u^2}
\left[\log(m_i^2(u)/\bar{\mu}^2)+\frac{1}{2}-C_i\right]
\label{m2UT}
\end{equation}
where, as explained in Appendix A, we shall take $\bar{\mu} = T$. In
the above, 
$i=g,g',\omega,\rho,f,{\cal H},{\cal Q}$ and $u\equiv\langle U \rangle
=\sqrt{-3 m_U^2/g_s^2}$.  

We will perform, as in the case of appendix A, 
resummation on the $n=0$ modes of
the longitudinal components of the gauge bosons $g_L,g'_L$ and for all modes of
the bosons $\omega,\rho$ and the light scalar ${\cal H}$. No resummation is
performed either on fermions $f$ or for the heavy scalar boson 
${\cal Q}$, whose mass
is dominated by the soft-term $m_Q$ and  is decoupled 
from the thermal bath.
We can then write the one-loop effective potential as in Eq.~(\ref{1loopot}),
where the sum extends over $i=g_L,g'_L,g_T.g'_T,f,\omega,\rho,{\cal H},{\cal
Q}$, and the masses $M_i$ are defined as:
\begin{equation}
M_i=
\left\{
\begin{array}{ll}
m_i & i=g_L,g'_L,g_T,g'_T,f,{\cal Q} \\ & \\
\overline{m}_i & i=\omega,\rho,{\cal H}
\end{array}\ .
\right.
\label{masasU}
\end{equation}
The thermal contributions $J^{(i)}$ are defined
\begin{equation}
J^{(i)}=
\left\{
\begin{array}{ll}
{\displaystyle J_B(m_i^2)-\frac{\pi}{6}\left(\overline{m}_i^3-m_i^3\right)} &
i=g_L,g'_L,\gamma_L \\ & \\
J_B(m_i^2) & i=g_T,g'_T,{\cal Q} \\ & \\
J_B(\overline{m}_i^2) & i=\omega,\rho,{\cal H} \\ & \\
J_F(m_i^2) & i=f \ .
\end{array}
\right.
\label{jotasU}
\end{equation}

The two loop diagrams that contribute to $V_2$ are, as in appendix A, 
of two kinds:
sunset diagrams labeled by $V_{XYZ}$, where $X$, $Y$ and $Z$ are
propagating fields, and figure eight diagrams labeled by $V_{XY}$, 
with propagating $X$ and $Y$ fields.
In the following we will denote: $G\equiv(g,g')$ and 
$\st\equiv(\omega,\rho)$, $\eta$ being the ghost fields. The relevant
contributions to $V_2$ are then (see also Ref. \cite{Schmidt}):
\begin{eqnarray}
V_{GGG} & = & -g_s^2\frac{N_c}{4}\left[ (N_c-2)\dvvv (\mg,\mg,0) +
\dvvv (\mg,\mg,\mgp)\right] \nonumber\\  
& \nonumber\\
V_{\eta\eta\, G} & = & -g_s^2\frac{N_c}{2} \left[ 2(N_c-1){\cal D}_{\eta\eta
V}(0,0,\mg)+{\cal D}_{\eta\eta V}(0,0,\mgp)\right] \nonumber \\
& \nonumber\\
V_{\st\st\, G} & = & -\frac{g_s^2}{4} \left[ (N_c-1)\dssv
(\momegabar,\momegabar,\mg)+(N_c-1)\dssv(\momegabar,\mrhobar,\mg)
\right. \nonumber\\
& + & 
\frac{N_c-1}{N_c}\dssv (\momegabar,\mrhobar,\mgp)
+\frac{1}{N_c}\dssv(\momegabar,\momegabar,\mgp)
\nonumber\\
&+ & \left. N_c(N_c-2)
\dssv(\momegabar,\momegabar,0)\right] \nonumber\\ & \nonumber \\
V_{\st\, GG} & = &-g_s^2\frac{\mg^2}{8}\left[
(N_c-1)\dsvv(\mrhobar,\mg,\mg)
+2\frac{(N_c-1)^2}{N_c^2}\dsvv(\mrhobar,\mgp,\mgp)
\right. \nonumber\\
& + & \left. \frac{(N_c-2)^2}{N_c}\dsvv(\momegabar,\mg,\mgp)+
N_c(N_c-2)\dsvv(\momegabar,\mg,0)\right] \nonumber\\ & \nonumber\\
V_{GG} & = & -g_s^2\frac{N_c}{8}\left[2(N_c-2)\dvv(0,\mg)+2\dvv(\mg,\mgp)+
(N_c-1)\dvv(\mg,\mg)\right]\nonumber \\ & \nonumber\\
V_{\st\, G} & = & -\frac{g_s^2}{8}\left\{ (N_c-1)\left[3\dsv(\momegabar,\mg)+
\dsv(\mrhobar,\mg)\right] \right. \nonumber\\
& + & \frac{1}{N_c}
\left[(N_c+1)\dsv(\momegabar,\mgp)+(N_c-1)\dsv(\mrhobar,\mgp)
\right] \nonumber\\
& + & \left. 2N_c(N_c-2)\dsv(\momegabar,0) \right\} \nonumber\\ & \nonumber\\
V_{\st\st\st} & = & -\frac{g_s^4}{18}\left[3H(\mrhobar,\mrhobar,\mrhobar)+
(2N_c-1)H(\mrhobar,\momegabar,\momegabar)\right]\, T^2\, U^2
\nonumber\\ & \nonumber\\
V_{\st{\cal H} {\cal H}} & = &  - 2 \left[ h_t^2\sin^2\beta\, U + h_t
\widetilde{A}_t\sin\beta \sin\overline{\theta}_{\cal H} 
\cos\overline{\theta}_{\cal H}\right]^2
H(\momegabar,\mhbar,\mhbar)\; T^2 \nonumber\\ & \nonumber\\
V_{\st\st} & = & \frac{1}{24}g_s^2\left[
3I^2(\mrhobar)+(4N_c-2)I(\mrhobar)I(\momegabar)+(4N_c^2-1)I^2(\momegabar)
\right] \nonumber\\ & \nonumber\\
V_{\st {\cal H}} & = & h_t^2 \sin^2\beta 
\cos^2\overline{\theta}_{\cal H}I(\mhbar)\left[
I(\mrhobar)+(2N_c-1)I(\momegabar)\right]
\label{2loopU}
\end{eqnarray}
where all functions involved in (\ref{2loopU}) are defined in 
Ref.~\cite{AE} and
cancellation of poles in $1/\epsilon$, and $\iota_\epsilon$ and $c_B$ terms, as
well as linear terms in $M_i T^3$, is achieved as in the case of appendix A
against counterterms and thermal counterterms.

\newpage
\begin{figure}[htb]
\centerline{
\psfig{figure=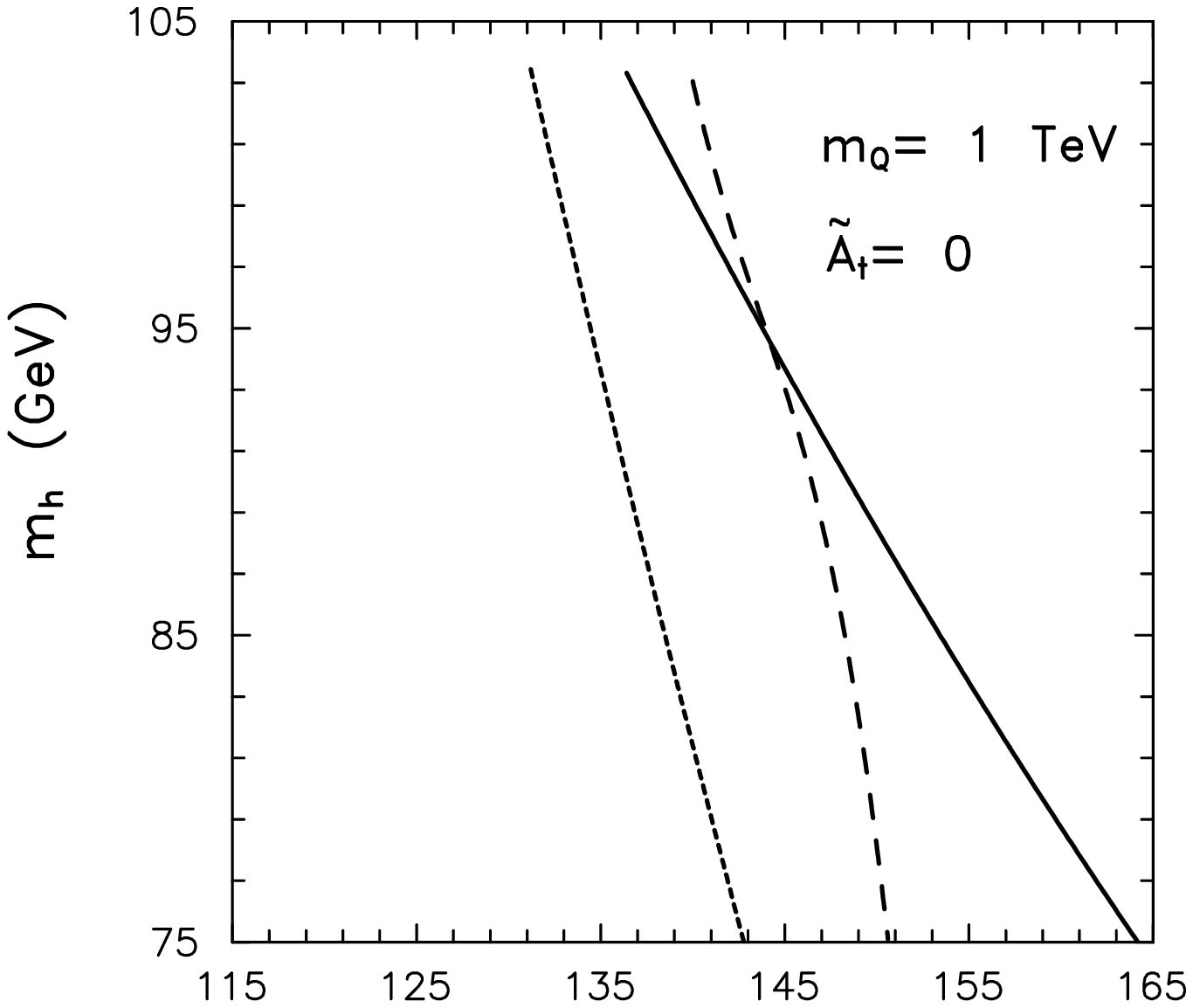,width=8.5cm,height=8.5cm,bbllx=5.cm,bblly=4.cm,bburx=
19.cm,bbury=17cm}
\psfig{figure=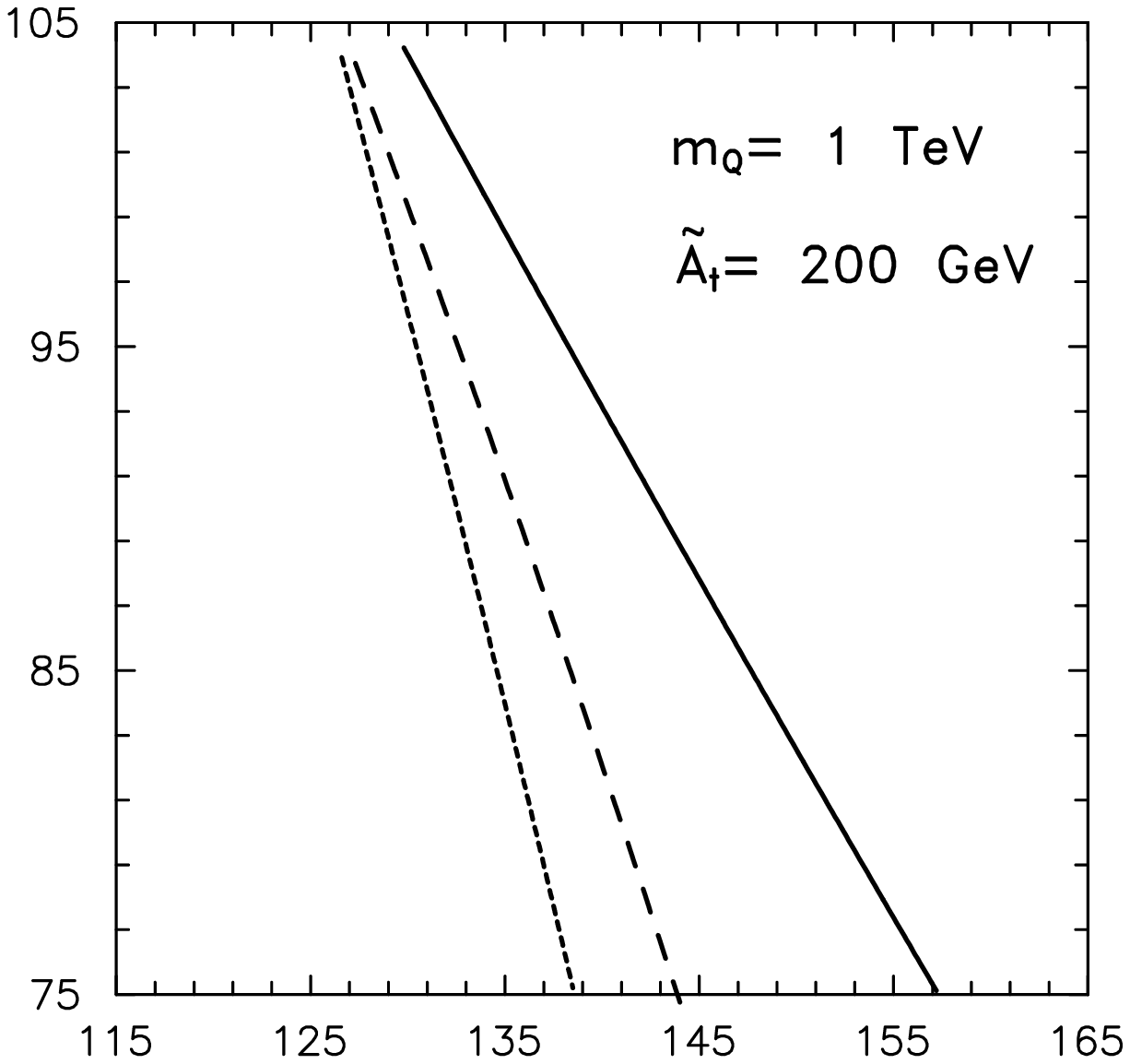,width=8.5cm,height=8.5cm,bbllx=5.cm,bblly=4.cm,bburx=
19.cm,bbury=17cm}}
\centerline{
\psfig{figure=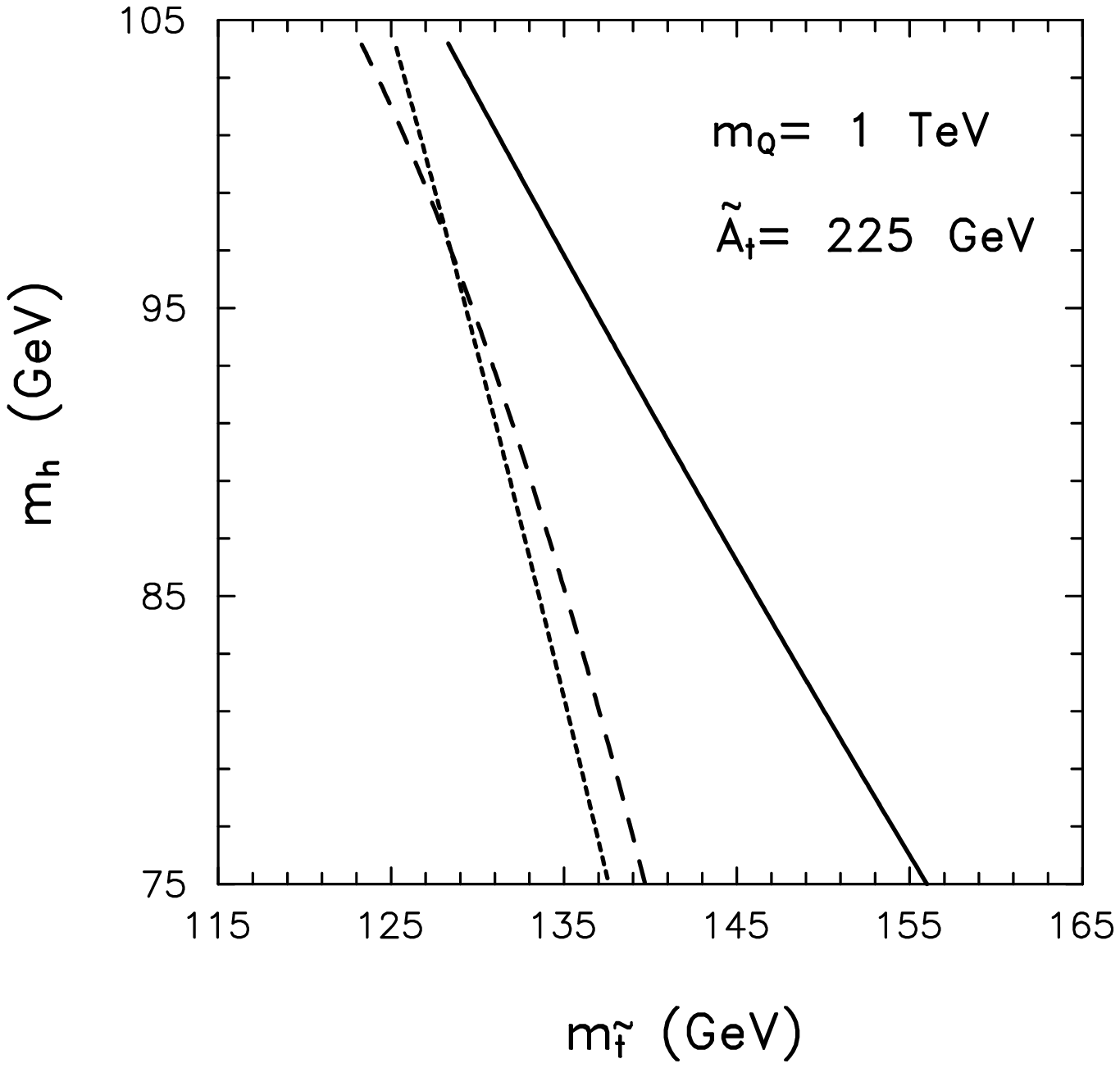,width=8.5cm,height=8.5cm,bbllx=5.cm,bblly=4.cm,bburx=
19.cm,bbury=17cm}
\psfig{figure=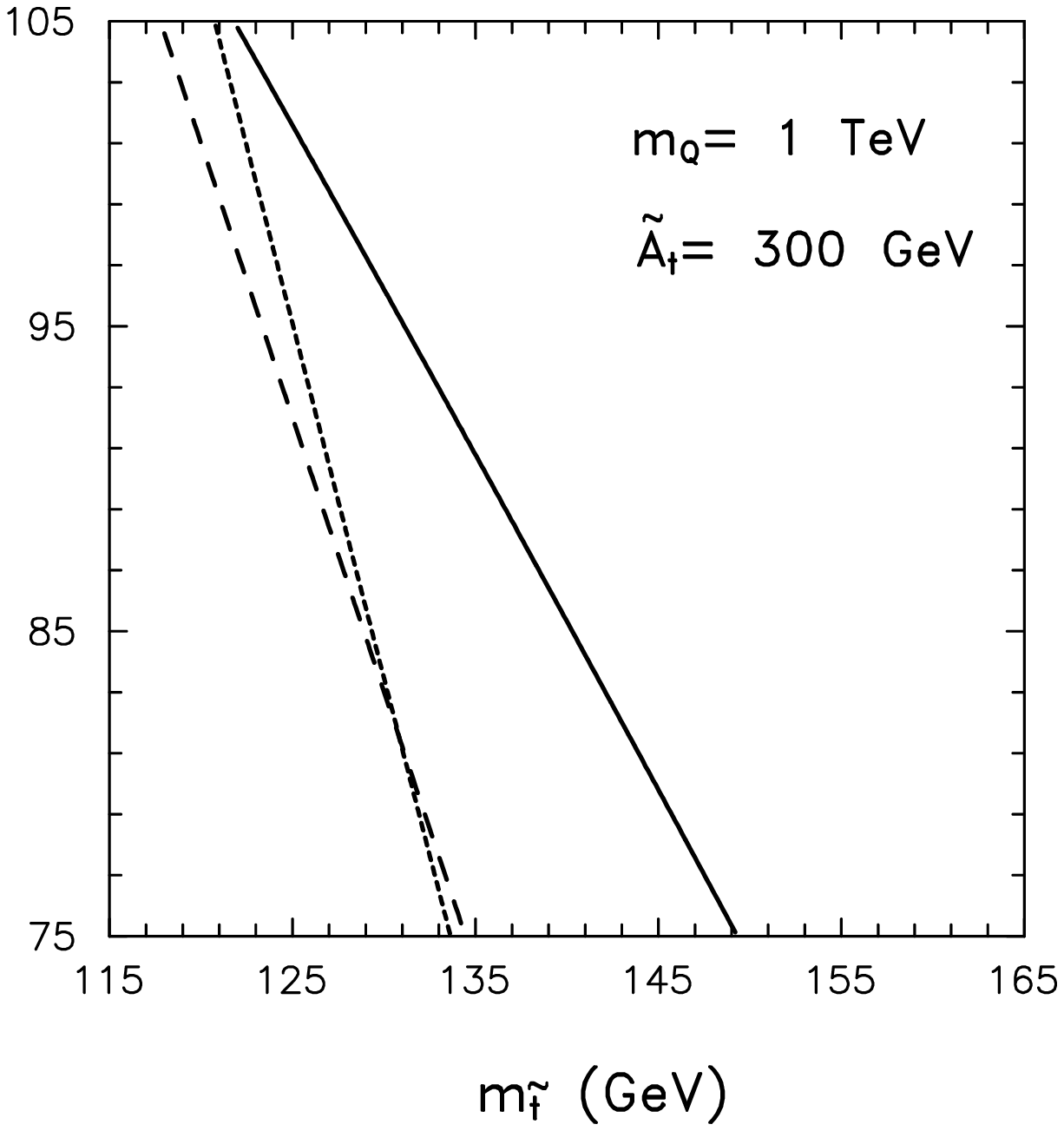,width=8.5cm,height=8.5cm,bbllx=5.cm,bblly=4.cm,bburx=
19.cm,bbury=17cm}}
\caption{Values of $m_h$, $m_{\st}$ for which $v(T_c)/T_c = 1$ (solid
line), $T_c^U = T_c$ (dashed line), $\widetilde{m}_U = 
\widetilde{m}_U^c$ (short-dashed line), for $m_Q = 1$ TeV
and different fixed values of  $\widetilde{A}_t$.
The region on the left of the solid line is consistent with a strongly
first order phase transition. A two step phase transition may occur in
the regions on the left of the
dashed line, while on the left of the short-dashed line, the physical
vacuum at $T = 0$
becomes metastable. The region on the left of both the dashed
and short-dashed lines leads to a stable color breaking vacuum state
at zero temperature and is hence physically unacceptable.}
\end{figure}
\begin{figure}[bht]
\centerline{
\psfig{figure=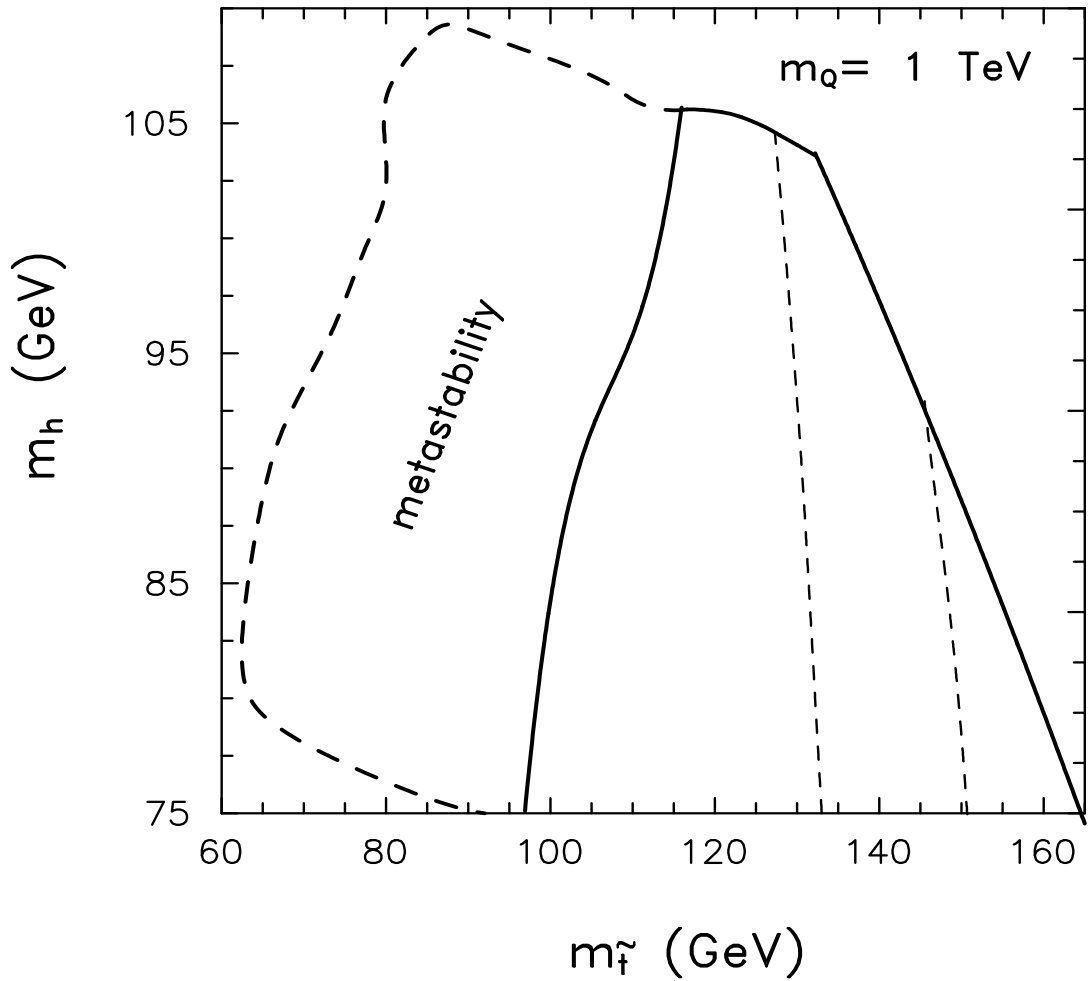,width=15.0cm,height=13.0cm,bbllx=5.cm,bblly=4.cm,bburx=
19.cm,bbury=17cm}}
\caption{Region of the $m_h$--$m_{\st}$ parameter space for
which a strongly first order phase transition takes place is
shown within solid lines. The short-dashed lines demark the
region for which a two-step phase transition may occur. The
region on the right of the dashed line and left of the short-dashed
may lead to a metastable vacuum state.}
\end{figure}
\end{document}